\documentclass[journal, 12pt]{IEEEtran}
\usepackage{cite}
\ifCLASSINFOpdf
  \usepackage[pdftex]{graphicx}
\else
  \usepackage[dvips]{graphicx}
\fi
\usepackage{array}
\usepackage[caption=false,labelformat=simple]{subfig}

\usepackage{stfloats}

\usepackage{multirow}
\usepackage{xcolor}

\begin{document}

%\title{Private NGSO Satellite Constellations for \\Future 6G Space-Terrestrial Networks}

\title{Towards Global and Limitless Connectivity:\\ The Role of Private NGSO Satellite Constellations for Future Space-Terrestrial Networks}

% Exploring Private NGSO Satellite Constellations for Designing Future 6G Space-Terrestrial Networks

\author{Andra M. Voicu, Abhipshito Bhattacharya, and Marina Petrova% <-this % stops a space
\thanks{Andra M. Voicu and Abhipshito Bhattacharya are with Mobile Communications and Computing, RWTH Aachen University, Aachen, Germany, E-mail: \mbox{voicu@mcc.rwth-aachen.de}, \mbox{bhattacharya@mcc.rwth-aachen.de}.}
%$\{voicu\vert bhattacharya\}$@mcc.rwth-aachen.de}
\thanks{M. Petrova is with Mobile Communications and Computing, RWTH Aachen University, Germany and KTH Royal Institute of Technology, Stockholm, Sweden, E-mail: \mbox{petrova@mcc.rwth-aachen.de}.}% <-this % stops a space
}

\maketitle

\begin{abstract}

Satellite networks are expected to support global connectivity and services via future integrated 6G space-terrestrial networks (STNs), as well as private non-geostationary satellite orbit (NGSO) constellations.
In the past few years, many such private constellations have been launched or are in planning, e.g. SpaceX and OneWeb to name a few. 
In this article we take a closer look at the private constellations and give a comprehensive overview of their features. We then discuss major technical challenges resulting from their design and briefly review the recent literature addressing these challenges. 
Studying the emerging private constellations gives us useful insights for engineering the future STNs. 
To this end, we study the satellite mobility and evaluate the impact of two handover strategies on the space-to-ground link performance of four real private NGSO constellations. 
We show that the link capacity, delay, and handover rate vary across the constellations, so the optimal handover strategy depends on the constellation design. Consequently, the communications solutions of future STNs should be compliant with the constellation specifics, and the STN standards need to be flexible enough to support satellite operation with the large parameter space observed in the emerging private constellations.

%We first summarize the state of the art of satellite communications and find that most works demonstrated the performance of their proposed algorithms for simple legacy or hypothetical constellations. 
%Second, we discuss the different design of the new private constellations and how this affects the different segments of the satellite network.
%Third, we present extensive simulation results for the space-to-ground links of the new NGSO private constellations in the Ka band. 
%We show that, in coexisting deployments, inter-constellation interference does not decrease the link throughput of each constellation equally and this depends on the constellation size, geometric properties, and transceiver parameters.
%Furthermore, in standalone constellations, the handover strategy has a marginal impact on the data rate. However, the preferred handover strategy is different for different constellations, when measured by the propagation delay and handover rate.        
%We thus argue that the design diversity of the new private NGSO constellations makes them a valuable and comprehensive source of real STN design parameters. Consequently, they should be considered when designing 6G STNs, regardless of whether coexisting in shared bands, or operating in standalone deployments.

\end{abstract}


% \fi
\IEEEpeerreviewmaketitle

%\section{Introduction}
%\IEEEPARstart{T}{his} demo file is 

\section{Introduction}

\IEEEPARstart{S}{atellite} constellations will play a crucial role in providing ubiquitous connectivity as an integral part of 5G-and-beyond wireless networks. They could support a variety of services such as mobile broadband and fixed Internet connectivity for ground users in unserved and underserved areas, passengers on board of airplanes, as well as wireless connectivity for Internet-of-Things (IoT), tracking ships and their cargos, backhaul for ground base stations (BSs) or unmanned aerial vehicles (UAVs)~\cite{Kodheli2021, Tataria2021}. 
Non-geostationary satellite orbit (NGSO) systems are especially suited for such applications. Being at lower altitudes compared to legacy geostationary satellite orbit (GSO) systems significantly reduces the communication delay  and allows NGSO systems to support higher data rates at a lower transmit power.
Consequently, the 3GPP standardization body aims at integrating NGSO satellite communications into terrestrial 5G and future 6G cellular networks~\cite{Maeaettaenen2019}. 

In recent years, driven by the commercial potential, numerous private companies outside 3GPP have also been investing in NGSO satellite communications systems and many of them have engineered and even started launching satellite constellations, e.g. SpaceX, OneWeb, Kepler, and O3b.
Hence, in addition to standardization efforts in 3GPP, several proprietary satellite communications solutions are being developed to provide connectivity to ground and airborne devices.

These new private NGSO constellations have very different characteristics in terms of orbit geometries, number of satellites, transmit power, and antennas, comprising low Earth orbit (LEO), medium Earth orbit (MEO), highly elliptical orbit (HEO), and geosynchronous constellations~\cite{ScheduleS2021, Braun2019, Tonkin2018}.
They are thus significantly more diverse than older-generation NGSO constellations like Iridium and Globalstar, which were small LEO constellations.  
The inherent high complexity and diversity of these NGSO satellite architectures combined with the high-speed mobility with respect to the Earth’s surface impose multiple resource management challenges that need to be carefully addressed. For instance, the high orbital speed will cause frequent handovers between satellites during the service time. This will also have implications on the load-balancing mechanisms that need to guarantee balanced traffic loads among satellite links in scenarios with spatially heterogeneous traffic demands.

The already planned private NGSO constellations will operate in the same frequency bands, i.e. the Ku- (10.7--12.7~GHz dowlink, 14.0--14.5 uplink), Ka- (17.8--19.7~GHz downlink, 27.5--30.0~GHz uplink), and V-band (37.5--42.0~GHz downlink, 47.2--51.4~GHz uplink)~\cite{Tonkin2018}. In these bands the regulation to protect one another from interference is rather loose, so interference mitigation, co-existence and inter-constellation spectrum sharing become key issues~\cite{Braun2019, Tonkin2018}. Thus, the performance of the existing spectrum sharing techniques for satellite systems needs to be revisited to ensure fair and effective operation in a rather dense and heterogeneous environment due to the emerging constellations.  
%It is with high probability that the future (standardized) space-terrestrial networks (STNs) will also target these bands, so this calls for a truly agile spectrum management solutions to solve the inter-satellite coexistence challenges. 

%they will have many design characteristics and communications challenges in common to the new private constellations. 

\begin{figure*}[t]
\centering	
\includegraphics [width = 0.75\linewidth] {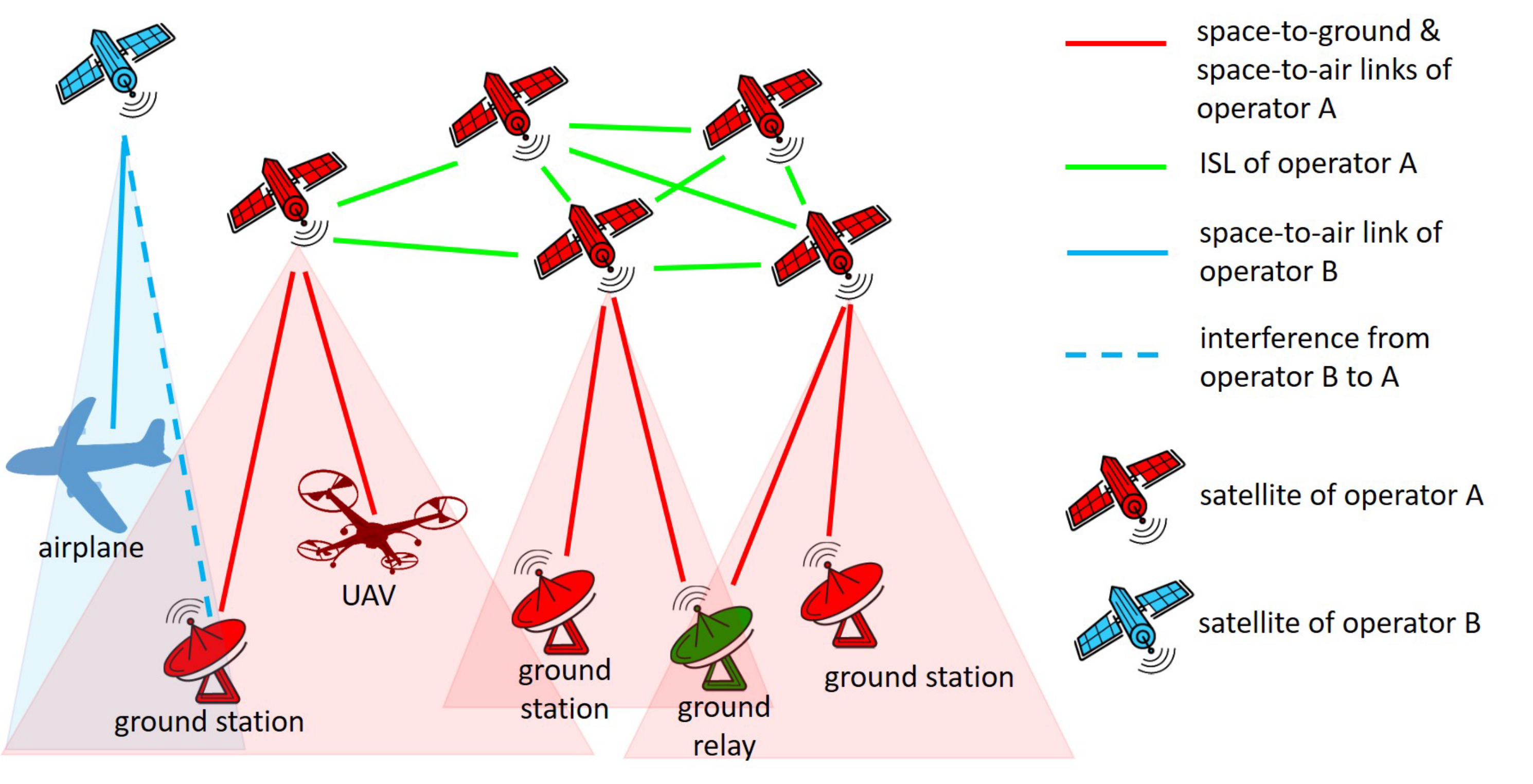}
\caption{Illustration of satellites managed by two operators, serving different ground stations, UAVs, and airplanes via space-to-ground and space-to-air links. Routing within the satellite network of operator A (red) occurs via ISLs or via ground relays. The satellite of operator B (blue) causes interference to the ground station served by operator A, which is also in the coverage area of operator B.}
\label{fig_concept}
\end{figure*}

%Thus, when designing new protocols and solutions for future STNs, we argue that the new private NGSO constellations are highly relevant due to the following reasons.
%The design diversity of these constellations, which covers a wide range of parameter configurations, is relevant for future 6G STNs from three perspectives. 
%First, the design of the private constellations is very diverse, so they are rich examples of real possible constellation configurations. Thus, they can be utilized as use cases when developing and verifying engineering solutions for future STNs. 
%Second, since it is expected that emerging private constellations and future STNs will coexist in shared bands, these private constellations should be considered for designing and tuning inter-constellation spectrum sharing mechanisms.  
%Third, for some cellular STN architectures, it would be possible to integrate private constellations for transmissions over the air interfaces, as non-3GPP interfaces~\cite{Guidotti2019}. Thus, future 6G STNs could benefit from using private constellations that will be deployed prior the finalization of the 6G standardization. 

There is no doubt that the new private NGSO constellations will play a key role in complementing the terrestrial networks and provide global high-speed and low-latency network connectivity. Nevertheless, several technical, regulatory, and business challenges need to be resolved on the way. In this article we give a comprehensive overview of the characteristics of the new private NGSO constellations and discuss the main technical challenges in different network segments, i.e. space-to-air, space-to-ground, and inter-satellite. 
These segments are illustrated in Fig.~\ref{fig_concept}, where two operators serve ground and airborne end devices. 
We then focus on the challenges caused by satellite mobility and, as far as we are aware, are the first to study the impact that different handover strategies (HO strategies) have on the space-to-ground capacity and delay in several real private NGSO constellations emerging in the Ka-band. 
We show that the optimal HO strategy depends on the constellation design and features. Consequently, we argue that the engineering solutions for future STNs should take into account the specific constellation design, e.g. by means of adaptive algorithms. Moreover, the future STN standard specifications should support the wide range of parameter values observed in emerging private NGSO constellations. 

The rest of the article is organized as follows.
We first present the major design features of the new private NGSO constellations in Section~\ref{sec_newNGSO}. 
In Section~\ref{sec_challenges} we discuss challenges resulting from these features and we briefly review how such challenges have been addressed in the recent literature. 
%We note that most works considered either legacy or hypothetical constellation configurations, typically with simple polar orbits, which are not sufficient to characterize the \textcolor{red}{currently deployed} NGSO systems.
Next, we shed light on the impact of HO strategies on the space-to-ground link capacity and delay in Section~\ref{sec_res_ho_capacity} and~\ref{sec_res_ho_delay}, respectively. 
%We thus demonstrate that the characteristics of the private NGSO satellite constellations can be used for learning valuable lessons for designing future STNs.
Finally, in Section~\ref{sec_conclusions} we conclude the article.

\begin{figure*}[t]
\centering	
\includegraphics [width = 0.85\linewidth] {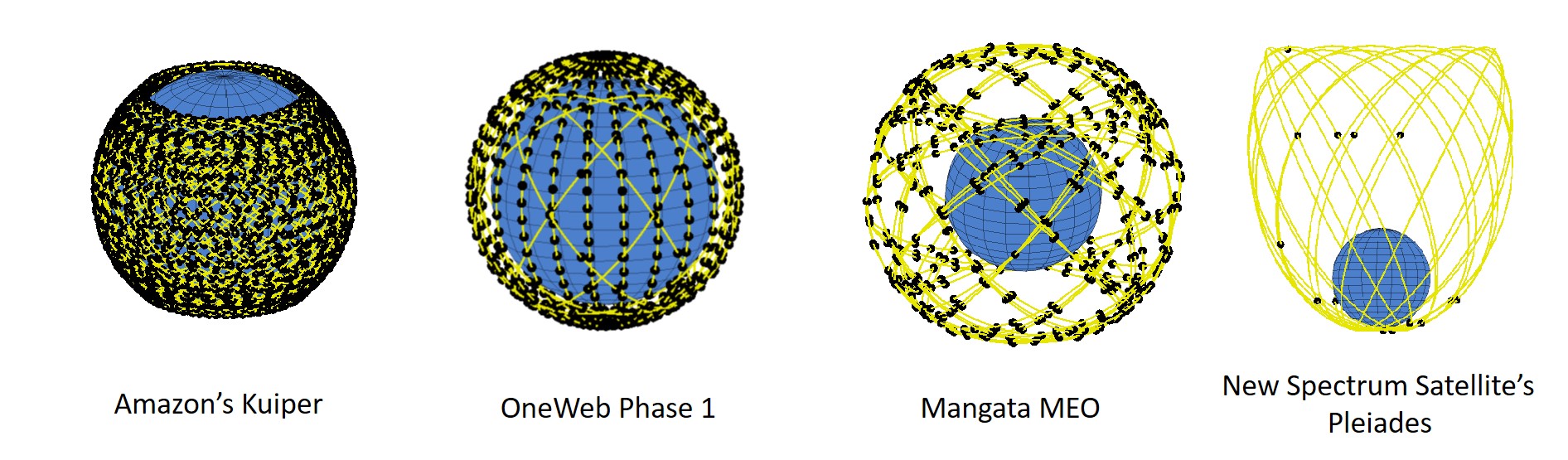}
\caption{Examples of new private NGSO satellite constellations in the Ka-band, showing the Earth (blue), the satellite orbits (yellow), and the satellites (black). Table~\ref{tab_const} summarizes the main design parameters of these constellations.}
\label{fig_constel}
\end{figure*} 

\begin{table*}[t!]
\centering
\caption{New private NGSO constellations and their main satellite parameters in the downlink~\cite{ScheduleS2021}}
\label{tab_const}
\begin{tabular}{|m{2.4cm}|m{2.1cm}|m{2cm}|m{1.3cm}|m{3cm}|m{1.8cm}|m{1cm}|m{1cm}|}
\hline
\centering \textbf{Constellation} 
	& \centering \textbf{Orbit type} 
	& \centering \textbf{Altitude [km]} 
	& \centering \textbf{No. satellites} 
	& \centering \textbf{Service \& feeder channel width [MHz]} 
	& \centering \textbf{EIRPD max. [dBW/Hz]} 
	& \centering \textbf{Band}
	& \centering \textbf{Started launch} \tabularnewline
	% & \centering \textbf{3 dB beamwidth} 
\hline
Kepler 
	& LEO (circular)
	& \center 650
	& \center 360
	& \center 10, 100, 300, 500 
	& \centering --41.0 
	& \centering Ku, Ka
	& \centering yes \tabularnewline
	% & \centering 3° 
\hline
Amazon's Kuiper 
	& LEO (circular)
	& \centering 590--630
	& \centering 3236
	& \centering 100 
	& \centering --43.9 
	& \centering Ka
	& \centering no \tabularnewline
	% & \centering 1.40° 
\hline
\multirow{3}{*}{Mangata} 
	& MEO (circular)
	& \centering 6400 
	& \centering 567
	& \centering 100, 500 
	& \centering --36.3 
	& \centering Ka, V
	& \centering \multirow{3}*{no} \tabularnewline
	% & \centering 0.25°
\cline{2-7}
    & HEO (elliptical)	
    & \centering perigee:~$>$1200 apogee:~$<$11600
    & \centering 224
    & \centering 100, 500
    & \centering --36.3 
    & \centering Ka, V
    & \tabularnewline
    % & \centering 0.25°
\hline
\multirow{2}{*}{O3b} 
	& LEO (circular)
	& \centering 507
	& \centering 36
	& \centering 250, 260, 300, 500, 2500 
	& \centering --22.5 
	& \centering Ka
	& \centering \multirow{2}*{yes} \tabularnewline
	% & \centering 3.5°
\cline{2-7}
	& MEO (circular)
	& \centering 8062
	& \centering 76
	& \centering 250, 260, 300, 500, 2500  
	& \centering --22.5 
	& \centering Ka
	& \tabularnewline
	% & \centering 3.5°	
\hline
OneWeb Phase 1 
	& LEO (circular)
	& \centering 1200
	& \centering 716
	& \centering 155, 250 
	& \centering --38.7
	& \centering Ku, Ka
	& \centering yes \tabularnewline
	% & \centering 0.16°
\hline
OneWeb Phase 2 
	& LEO (circular)
	& \centering 1200
	& \centering 47844
	& \centering 155, 250 
	& \centering --38.7 
	& \centering Ku, Ka
	& \centering no \tabularnewline
\hline
SpaceX
	& LEO (circular)
	& \centering 540--570
	& \centering 4408 
	& \centering 50
	& \centering --50.3 
	& \centering Ku, Ka
	& \centering yes \tabularnewline
	% & \centering 0.8°  
\hline
SpaceX Gen2
	& LEO (circular)
	& \centering 328--614
	& \centering 30000 
	& \centering 50, 100, 500, 800, 2000
	& \centering --37.5 
	& \centering Ku, Ka
	& \centering no \tabularnewline
	% & \centering 0.8° 
\hline
Telesat
	& LEO (circular)
	& \centering 1000--1325
	& \centering 1788
	& \centering 500, 800 
	& \centering --50.0 
	& \centering Ka
	& \centering no \tabularnewline
	% & \centering 0.8° 
\hline
Viasat
	& LEO (circular)
	& \centering 1300
	& \centering 288
	& \centering 500, 800 
	& \centering --31.7 
	& \centering Ka, V
	& \centering no \tabularnewline
	% & \centering 2.0°
\hline
Karousel
	& geosynchronous (elliptical)
	& \centering perigee:~31569 apogee:~40002
	& \centering 12
	& \centering 250
	& \centering --22.7 
	& \centering Ku, Ka
	& \centering no \tabularnewline
	% & \centering 2.0° 
\hline
New Spectrum Satellite's Pleiades
	& HEO (elliptical)
	& \centering perigee:~1125 apogee:~26679
	& \centering 15
	& \centering 20, 25 
	& \centering --24.7 
	& \centering Ku, Ka
	& \centering no \tabularnewline
	% & \centering 2.0° 
\hline
Space Norway
	& HEO (elliptical)
	& \centering perigee:~8089 apogee:~43509
	& \centering 2
	& \centering 115, 250, 500, 1000 
	& \centering --26.0 
	& \centering Ku, Ka
	& \centering no \tabularnewline
	% & \centering 3.0° 
\hline
Theia
	& LEO (elliptical)
	& \centering perigee:~750 apogee:~809
	& \centering 120
	& \centering 1, 300, 400, 500, 1500 
	& \centering --43.5 
	& \centering Ku, Ka
	& \centering no \tabularnewline
	% & \centering 2.0° 
\hline
AST\&Science's SpaceMobile
    & LEO (elliptical)
    & \centering perigee:~$>$725 apogee:~$<$740
    & \centering 243
    & \centering 500, 4500
    & \centering -36.8
    & \centering V
    & \centering no \tabularnewline
\hline
\multirow{3}{*}{Boeing} 
	& LEO (circular)
	& \centering 1056 
	& \centering 132
	& \centering 2000, 2500 
	& \centering --1.8
	& \centering V
	& \centering \multirow{3}*{no} \tabularnewline
	% & \centering 0.25°
\cline{2-7}
    & HEO (elliptical)	
    & \centering perigee:~$>$27354 apogee:~$<$44222
    & \centering 15
    & \centering 2000, 2500
    & \centering --1.8 
    & \centering V
    & \tabularnewline
\hline
\end{tabular}
\end{table*}

\section{Characteristics \& Design Parameters of New Private NGSO Constellations}
\label{sec_newNGSO}

The emerging private NGSO constellations have very different design parameters and this can be best observed in the formal applications submitted by the satellite operators to the US spectrum regulator FCC in the last few years~\cite{ScheduleS2021}. 
%Moreover, most of them are very different than the legacy constellations like Iridium and Globalstar, which have been largely considered in the literature thus far, \emph{cf.} Section~\ref{sec_litrev}.
%We focus on the constellations that have been proposed to operate in the Ka band and for which formal applications have been submitted to the FCC, i.e. the US spectrum regulator. We select the NGSO constellations in the Ka band, since numerous satellite  operators are currently targeting this band and the proposed NGSO constellations are more diverse than in the Ku and V bands, which are also being considered for operation of some new NGSO constellations~\cite{ScheduleS2021}. 
Table~\ref{tab_const} summarizes some main design features of the new constellations proposed to operate in the Ku-, Ka-, and V-band and Fig.~\ref{fig_constel} illustrates a few examples from the Ka-band.

From Table~\ref{tab_const}, it can been easily seen that the constellation size, i.e. number of comprised satellites, varies significantly. Mega-constellations like SpaceX are representative of the upper size range and include thousands and even tens of thousands (SpaceX Gen2) of satellites, while the lower range is represented by Space Norway with only two satellites. 

The new private NGSO constellations also have very diverse geometric orbit properties, such as inclination, altitude, and orbit shape. For instance, LEO satellites typically have a constant altitude that can be as low as 328~km (for SpaceX Gen2), whereas for the HEO constellations the altitude of the satellites varies over a few thousands of kilometers (e.g. Mangata HEO).
Moreover, the different altitudes and orbit shapes result in different velocities of the satellites, to keep them in their orbits.

Finally, although these NGSO constellations share the same spectrum bands, each satellite operator selects their own transmit power and antenna type (resulting in an individual effective isotropic radiated power density---EIRPD), as well as channels and channel widths for different purposes like service, feeder, and telemetry, tracking, and control (TT\&C), as evident in Table~\ref{tab_const}.  This calls for a design of effective co-existence and interference mitigation techniques, discussed in the next section.

\section{Challenges for Emerging NGSO Satellite Communication Systems}
\label{sec_challenges}

%Satellite communications pose similar challenges to both new NGSO constellations and other future STNs, regardless of them being standardized or private. This is 
Due to the intrinsic nature of the satellite systems, e.g. long communication distances, high mobility, and pre-defined constellation geometric properties, a number of challenges need to be resolved so that both private NGSO and the future STNs could offer global high-speed and low-latency Internet connections.
In the following, we point to these challenges, and briefly review how they have been addressed in the recent literature. 

%In this section we discuss such important challenges, some of which have also been identified for integrating STNs into 3GPP networks of different generations, i.e. LTE~\cite{Guidotti2019} and 5G~\cite{Maeaettaenen2019}. Furthermore, we briefly review how such challenges have been addressed in the more general literature on satellite communications and also for private NGSO constellations thus far.  

\paragraph{Capacity of Space-to-Ground \& Space-to-Air Links}

The emerging satellite systems are expected to offer high space-to-ground and space-to-air link data rates of 5~Gbps and higher~\cite{Kodheli2021}. However, in order to support a large number of devices per satellite at a certain quality of service (QoS), efficient load balancing and resource management schemes need to be in place. This is critical especially since the satellites cover a large area due to their high altitude, so many ground and airborne devices would connect to the same satellite. One possibility to manage this is through multiple-input-multiple-output (MIMO) antenna configurations, such that different end devices connect to different beams. For this, interference among the beams of the same satellite must be mitigated and this could be achieved by e.g. reusing the same frequency resources together with PHY layer precoding~\cite{You2020}. Thus, the capacity of a satellite would in principle increase proportionally with the number of supported beams. Nonetheless, this technique mitigates interference efficiently only if the channels of the beams are uncorrelated; otherwise, different time or frequency resources must be allocated to different beams.
Furthermore, accurately beamforming at the satellites is non-trivial and requires calibration via signals sent to a ground station~\cite{An2020}. 

Another possibility to increase the capacity in the area covered by a satellite is to use jointly space, air, and ground links. Such a solution was considered in~\cite{Hofmann2020}, where Internet connectivity for users on board of airplanes was ensured via a combined satellite and terrestrial 5G network, as well as inter-airplane links able to relay traffic. 
Finally, the Ku-, Ka-, and V-band, where the new private constellations will operate, are rather wide, so the per-satellite capacity of these constellations can be increased by simply allocating multiple frequency channels per satellite, as long as interference among adjacent satellites remains within reasonable limits.

%Previous works that have considered capacity issues for space-to-ground and space-to-air links addressed e.g. Internet connectivity on board of airplanes by means of a combined terrestrial 5G network, satellites, and inter-airplane links~\cite{Hofmann2020}, as well as antenna techniques to enable efficient beamforming at the satellites and thus increase the capacity, e.g.~\cite{You2020, LaMar2020, An2020}.
%Nonetheless, the satellite networks were represented in a simplistic way, where~\cite{Hofmann2020, You2020} considered a single satellite, \cite{LaMar2020} considered a single link with one transmitter and one receiver, while~\cite{An2020} presented no validation results at all. 
%Thus, the efficiency of the proposed engineering solutions to increase capacity has not yet been studied for realistic NGSO parameter configurations, as those of the private constellations in Section~\ref{sec_newNGSO}, which are also expected to be representative of future 6G STNs. 

\paragraph{Interference Among Constellations}

The private NGSO constellations are set to share the Ku-, Ka-, and V-band among each other, which could cause significant inter-constellation interference for the space-to-ground and space-to-air links. This is due to the large number of coexisting constellations, the large size of some constellations, and the different transceiver parameters, which turn some constellations into stronger interferers than others~\cite{Braun2019, Tonkin2018}. Moreover, since future STNs will likely transmit in the same bands, the interference is expected to increase even further, emphasizing the need to design efficient interference mitigation techniques for such cases. 

Some examples of interference mitigation techniques considered for satellite communications are \emph{band-splitting} among constellations if the interference exceeds a given threshold and \emph{look-aside}, namely selecting a useful link separated by at least a given angle from an interfering link. However, such techniques were designed in the context of few, legacy small constellations and it has been shown that they are not efficient if all the new private NGSO constellations operate at the same time over the same portion of spectrum~\cite{Braun2019, Tonkin2018}. Specifically, band-splitting results in a high throughput degradation with a median of even 83\%, due to the decrease in the bandwidth allocated per constellation. Furthermore, look-aside is sometimes beneficial for large constellations like SpaceX, but harmful for smaller constellations like Kepler, for which it degraded the throughput by even 24\% compared to the case of simply suffering from interference.  
Consequently, designing efficient interference mitigation techniques for emerging private constellations and other future STNs is still an open challenge.

%We note that the impact of interference among NGSO constellations has largely not yet been considered in the specific context of cellular networks and future 6G STNs.

%The authors in~\cite{Tonkin2018, Braun2019} studied interference and mitigation techniques for private NGSO constellations in the Ku, Ka, and V bands, but especially in the Ka band even more new private constellations have been proposed very recently, e.g. Kuiper and Mangata. In Section~\ref{sec_res_interf} we adopt the methodology in~\cite{Tonkin2018, Braun2019} and we show the impact of interference and the look-aside mitigation technique on the recent constellation configurations.  

\paragraph{Propagation Delay}

The long propagation delay is a common issue in satellite communications, due to the large distances over which the signals travel. 
This is especially challenging when integrating satellite systems into existing terrestrial networks, which are designed to tolerate much shorter delays.  
For instance, for integrating STNs into 3GPP networks, the satellite-to-ground round-trip delay of tens or even hundreds of milliseconds was found to affect several important procedures designed for below 1~ms terrestrial delays, e.g. the timing advance, random access, hybrid automatic repeat request (HARQ), and handovers~\cite{Guidotti2019, Maeaettaenen2019}. Thus, major standard modifications would be required to integrate STNs into future cellular networks like 6G. 
In Section~\ref{sec_res_ho_delay} we show the impact of handovers on the propagation delay over the space-to-ground links, for different new private NGSO constellations proposed to operate in the Ka-band.

\paragraph{Doppler Shift}

The high velocities of the NGSO satellites relative to the served end devices result in high Doppler shift values, which have to be compensated for.
Once again, this poses issues especially when integrating STNs into existing terrestrial networks like 3GPP cellular ones. Specifically, the Doppler shift for satellite-to-ground links can reach hundreds of kHz, unlike in e.g. LTE systems designed to tolerate shifts below 1~kHz. This significantly shifts the OFDM subcarriers and requires compensation solutions beyond those implemented for LTE or 5G terrestrial networks~\cite{Guidotti2019, Maeaettaenen2019}.

\paragraph{Handovers}
Due to the NGSO satellite mobility with different velocities, handovers are required for the satellite-to-ground links, in order to ensure continuous connectivity to the end devices. 
One major challenge for a HO strategy is to achieve a good tradeoff between the overhead associated with the number of performed handovers and the channel quality.
A few HO strategies have been proposed and the most popular examples include selecting the satellite with (i)~the longest service time, (ii)~the highest elevation angle (or equivalently, the shortest link distance), and (iii)~the highest number of free channels~\cite{Hu2018}.
When aiming to reduce the handover overhead, it is important to note that the satellites move according to predefined patterns, so handovers can be predicted and selected in a sequence that keeps the overhead low. This was addressed in~\cite{Hu2018} by means of graph theory, for different HO strategies. 
In Sections~\ref{sec_res_ho_capacity} and~\ref{sec_res_ho_delay} we study the first two HO strategies from the above list, namely (i) and (ii) and present their impact on the link capacity and propagation delay for four selected private NGSO constellations.

It is worth noting that in the context of different 3GPP cellular STN architectures, the need for handovers in the entire communication path between a user and the core network has to be taken into account. Specifically, for a pure terrestrial network, handovers are required only for the radio access link between a BS and an end device. By contrast, for a satellite acting as a BS, both the access link to the end device and the backhaul link to the core network are mobile and require handovers~\cite{Guidotti2019}.

%From the perspective of satellite communications in general, the authors in~\cite{Hu2018} considered handovers and focused on predicting the required handovers for the space-to-ground links, due to the mobility of both LEO satellites and ground users. Three major handover strategies were considered, i.e. selecting the satellite with (i)~the longest service time, (ii)~the highest elevation angle (or equivalently, the shortest link distance), and (iii)~the highest number of free channels. 
%However, performance results were presented only for the legacy Globalstar constellation, which comprised only 56 LEO satellites.

\paragraph{Routing}

Routing in satellite networks is challenging due to several reasons. 
First, due to the satellite mobility, the inter-satellite links (ISLs) are rendered temporary, so these links break and form very dynamically and so do the routes traversing them. Optimum ISL formation was addressed in~\cite{LeyvaMayorga2021}, but only for polar constellations, where rather stable links can be formed between parallel polar planes. It is likely that ISLs are formed in an even more dynamic way for orbits other than polar, as observed for many of the new private NGSO constellations in Fig.~\ref{fig_constel}.

Second, less sophisticated satellites do not support ISLs and route traffic via ground relays instead. This could be an issue since it was shown that, for SpaceX, this results in round-trip delays of even 20~ms longer than for routing via ISLs~\cite{Handley2019}. 
Third, the computational complexity for running routing algorithms for satellite networks can be rather high. 
In order to better control route computation, \cite{Papa2020} considered a software-defined network (SDN), where only a few dedicated satellites act as controllers and calculate routes.
Finally, the IP node addressing scheme used for routing in terrestrial networks is not well suited for routing via satellites, since the IP address of a satellite moves over different geographical regions and is not permanently associated to a specific area. This was considered by means of a new addressing scheme in~\cite{Roth2021}.

\begin{figure*}[t]
\centering	
\subfloat[link spectral efficiency]{\includegraphics [width = 0.95\linewidth] {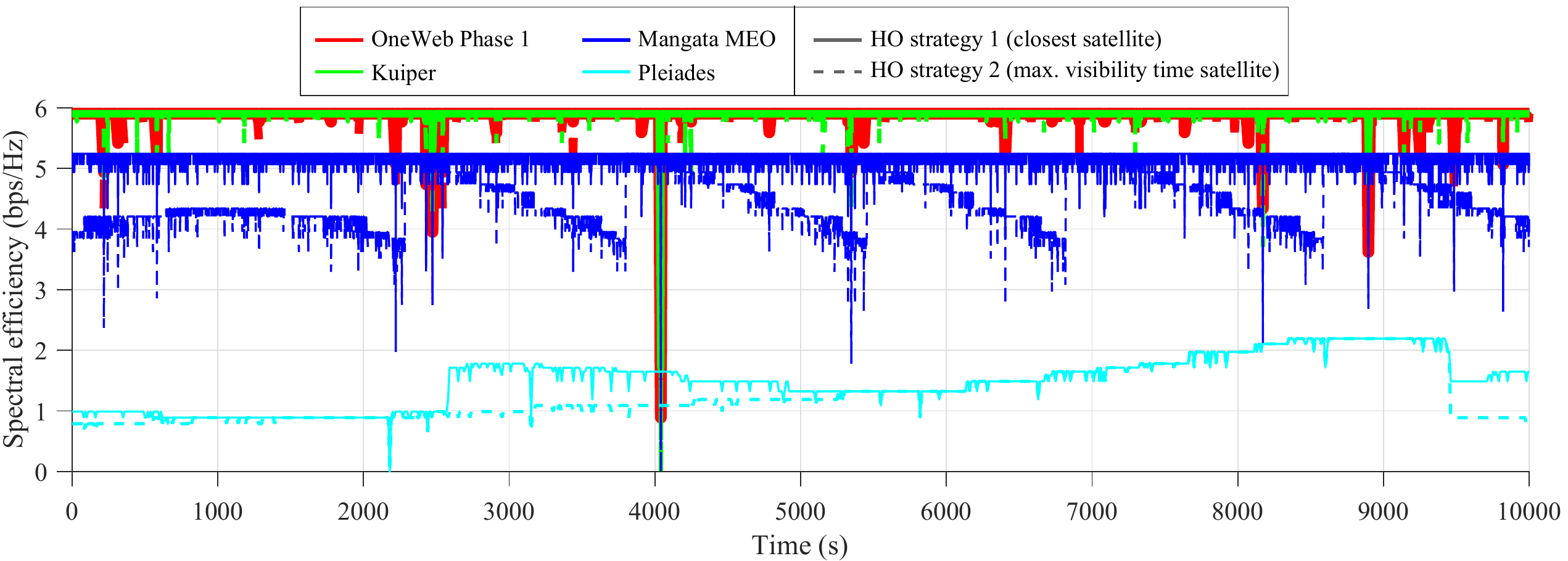}\label{fig_ho_cap_1}}
\\
\subfloat[link data rate]{\includegraphics [width = 0.95\linewidth] {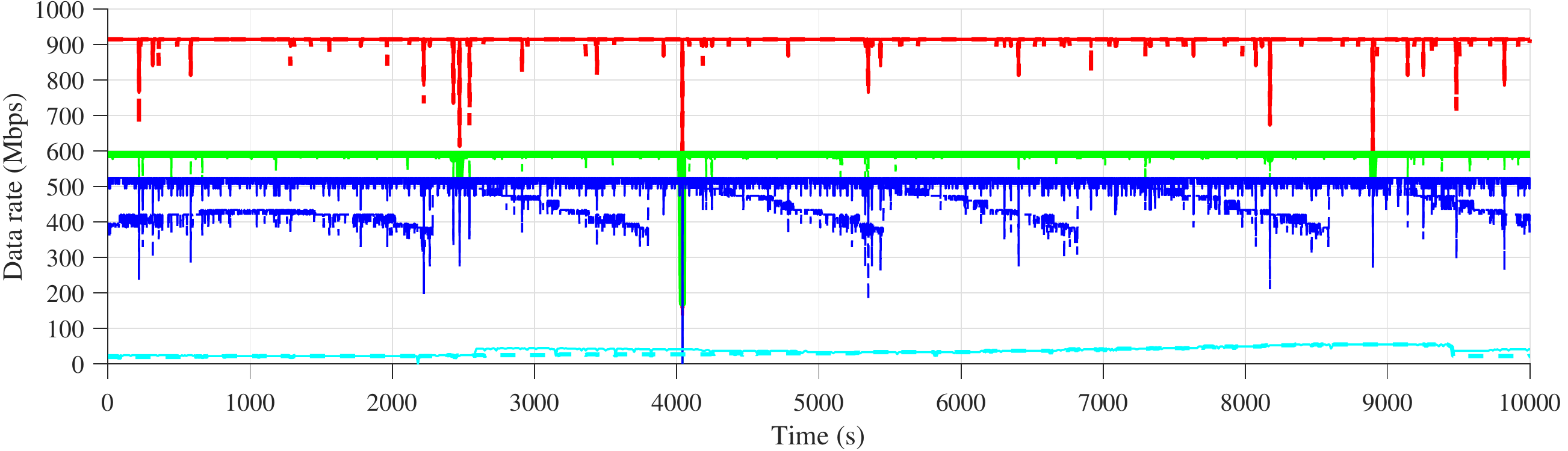}\label{fig_ho_cap_2}}
\caption{Link spectral efficiency and data rate for the two HO strategies, for different private NGSO constellations and a ground station located in Aachen, Germany. The used channel widths are 155~MHz, 100~MHz, 100~MHz, and 25~MHz for OneWeb Phase 1, Kuiper, Mangata MEO, and Pleiades, respectively.}
\label{fig_ho_cap}
\end{figure*}

\section{Impact of the HO Strategies on the Space-to-Ground Link Capacity}
\label{sec_res_ho_capacity}

As discussed in Section~\ref{sec_challenges}, handovers are critical procedures in STNs, since they handle the intrinsic satellite mobility and ensure that the end devices are always connected. 
Different HO strategies result in a different performance of the space-to-ground and space-to-air links, depending on when a handover is performed and which new satellite an end device is handed over to.
In this section we consider the example of a space-to-ground link for a ground station located in Aachen, Germany and we evaluate the impact of two major HO strategies on the capacity of this link, for different real private NGSO constellations. 

The two selected HO strategies are~\cite{Hu2018}: \textbf{(i)}~\emph{HO strategy~1}, where the ground station always connects to the closest satellite; and \textbf{(ii)}~\emph{HO strategy~2}, where the ground station connects to the satellite with the maximum remaining visibility time and remains connected to this satellite until it goes out of visibility.
Thus, HO strategy~1 aims at minimizing the path loss and the propagation delay, but on the other hand maximizes the number of handovers and thus the associated overhead. By contrast, HO strategy~2 aims to minimize the number of handovers, but may lead to higher path losses and propagation delays. 

We evaluate the impact of these HO strategies on the space-to-ground link of four example constellations, i.e. Kuiper, OneWeb Phase~1, Mangata MEO, and Pleiades, with very different design properties, \emph{cf.} Fig.~\ref{fig_constel} and Table~\ref{tab_const}. We assume transmissions in the Ka-band and the corresponding transmit power and antenna parameters of these constellations specified for this band~\cite{ScheduleS2021}.
The constellations operate in a standalone mode where they do not interfere with each other and the ground station can connect to satellites in only one given constellation. 
We conducted extensive simulations for a total simulated duration of 10,000~s, using the MATLAB satellite simulator in~\cite{Tonkin2018, Braun2019} which we modified to capture consecutive discrete moments in time and incorporate the two considered HO strategies.

The space-to-ground link capacity is first estimated in terms of downlink \emph{spectral efficiency}, by mapping the downlink signal-to-noise ratio (SNR) to the spectral efficiency of DVB-S2X~\cite{Tonkin2018, Braun2019}. Furthermore, we estimate the capacity as a \emph{link data rate}, by multiplying the spectral efficiency by the channel widths of 155~MHz, 100~MHz, 100~MHz, and 25~MHz for OneWeb, Kuiper, Mangata MEO, and Pleiades, respectively.
The corresponding impact of the handovers on the \emph{propagation delay} is presented subsequently in Section~\ref{sec_res_ho_delay}.   

%We assume a subset of the private constellations in Table~\ref{tab_const} operating in the Ka band in the downlink and a ground station in Aachen, Germany. We quantify the performance of the HO strategies in terms of channel data rate and one-way propagation delay. We note that the  channel data rate is obtained based on the spectral efficiency, were we 

\begin{figure}[t]
\centering	
\includegraphics [width = 0.95\linewidth] {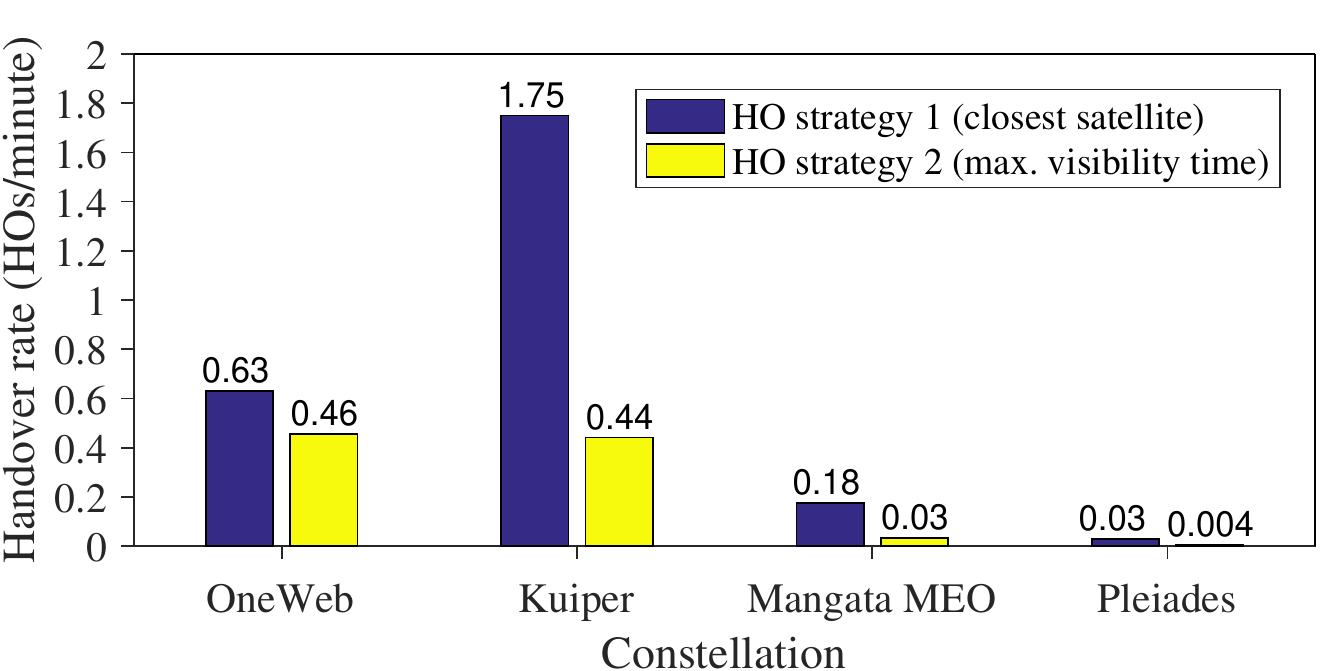}
\caption{Average handover rate for the two HO strategies, for different private NGSO constellations.}
\label{fig_ho_rate}
\end{figure}

Fig.~\ref{fig_ho_cap} shows the link spectral efficiency and data rate for the considered private constellations, for the two HO strategies, over the simulated time.
The spectral efficiency of the two LEO constellations (i.e. OneWeb and Kuiper) in Fig.~\ref{fig_ho_cap_1} is overall constant and reaches the maximum of 6~bps/Hz, regardless of the HO strategy. Consequently, for these constellations the choice of HO strategy does not affect the spectral efficiency. Thus, when selecting HO strategies for such LEO constellations, other metrics than the data rate should be in focus. For instance, we observe in Fig.~\ref{fig_ho_rate} that the number of handovers for Kuiper is significantly decreased by HO strategy~2 versus strategy~1, so strategy~2 would be preferred for reducing the control overhead, without trading off the link capacity.

\begin{figure*}[t]
\centering	
\includegraphics [width = 0.95\linewidth] {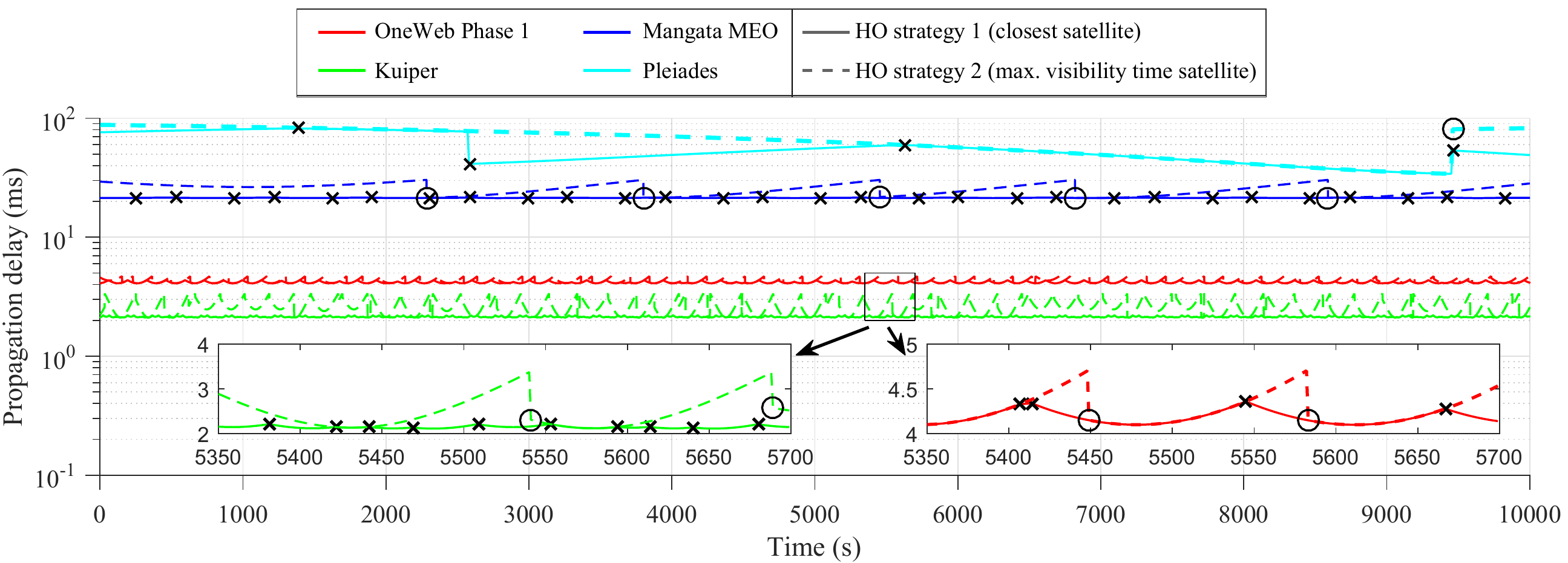}
\caption{Propagation delay for the two HO strategies, for different private NGSO constellations. The time instances when handovers are performed are marked for the first HO strategy ($\times$) and the second HO strategy ($\bigcirc$).}
\label{fig_ho_delay}
\end{figure*}

By contrast, for Mangata MEO and Pleiades HEO, HO strategy~1 achieves a somewhat higher spectral efficiency than HO strategy~2. This is since the EIRPD of these two constellations is not high enough to fully compensate for the path loss corresponding to large space-to-ground distances. Consequently, for these constellations, always selecting the closest satellite with strategy~1 is preferred for obtaining a high spectral efficiency.
Furthermore, the handover rate of these two constellations in Fig.~\ref{fig_ho_rate} is rather low for both HO strategies, so none of the strategies is preferred over the other from the perspective of the handover overhead. This shows that the impact of the HO strategies is different for LEO constellations versus MEO and HEO constellations.

The corresponding link data rate in Fig.~\ref{fig_ho_cap_2} varies over a wide range across the constellations (i.e. from 25~Mbps to 950~Mbps), corresponding to the different channels widths of 25--155~MHz. This is larger than the data rate difference of typically 100~Mbps observed between the two HO strategies for Mangata MEO. 
As such, supporting diverse channel widths can have a much stronger impact on the resulting link capacity than the handover choice. Consequently, other metrics like the delay should also be considered alongside capacity when designing handover algorithms for STNs.

Finally, the wide ranges of link capacity and handover rates observed for these different private NGSO constellations illustrate that the communication performance and operation modes of the NGSO satellite constellations can be very different. Consequently, we argue that future STN standards should be flexible enough to support all these operation ranges. 

%Furthermore, the data rate of a given constellation is virtually constant over the simulated time, for most constellations, i.e. OneWeb LEO, Kuiper, and Mangata MEO, regardless of the selected HO strategy. This shows that the transceiver parameters are well tuned to compensate for the variation of the path loss due to the different space-to-ground link distances within the constellation. 

%As an exception, for Pleiades the data rate is not constant over the simulated time, due to the poorly tuned transceiver parameters, as discussed for Fig.~\ref{fig_thrDeg}. Nonetheless, the data rate is only marginally different for the two HO strategies, so the choice of HO strategy does not significantly affect the data rate, consistent with the results for the other constellations. 

\section{Impact of the HO Strategies on the Space-to-Ground Link Propagation Delay}
\label{sec_res_ho_delay}

In this section we illustrate and discuss the impact of the two selected HO strategies in Section~\ref{sec_res_ho_capacity} on the propagation delay\footnote{In the remainder of this article the terms \emph{delay} and \emph{propagation delay} are used interchangeably.} of the space-to-ground links.
We assume the same four NGSO constellations and we show the corresponding one-way propagation delay and the instances when handovers are triggered in Fig.~\ref{fig_ho_delay}. 

The propagation delay varies significantly across the different constellations (i.e. from 2 to 90~ms), as expected due to their different altitudes.
Importantly, for each given constellation, the selection of the HO strategy has an impact on the delay. Specifically, the delay for HO strategy~1 (always closest satellite) is lower than for HO strategy~2 (staying connected to the satellite with maximum visibility time), as expected.
Although this effect is consistent across constellations, the specific difference in the propagation delay depends on the size and geometry of the constellations. 
For Kuiper, which is a very large LEO constellation, the delay is at most 1.5~ms longer for HO strategy~2 compared to strategy~1. This suggests that HO strategy~2 may be preferred for large LEO constellations, since the delay is not significantly longer, while the handover overhead is significantly lower, as discussed in Section~\ref{sec_res_ho_capacity}. 

By contrast, for Pleiades, the delay for HO strategy~2 is up to 25~ms longer than for HO strategy~1, while the handover rate of strategy~1 is very low and its capacity higher, \emph{cf.} Section~\ref{sec_res_ho_capacity}. A similar trend can be observed for Mangata MEO. As such, for less dense HEO and MEO constellations it is preferable to always select the closest satellite with HO strategy~1.  

Importantly, these results show overall that the optimal HO strategy depends on the constellation design, where strategy~1 is suitable for sparse MEO and HEO constellations, whereas strategy~2 benefits large LEO constellations.  
Such insights are useful for STNs in general and indicate that the engineering solutions for future STNs would have to be tuned according to the constellation design, potentially via adaptive algorithms.
Finally, as we observed very different delays for the different private constellations, we argue that future STN standards should support all these operation ranges, in order to achieve truly globally applicable specifications and inter-working compatibility with terrestrial deployments. 

%The results in this section show overall that the HO strategy has a significant impact on the delay and HO overhead, where for dense LEO constellations it is preferable that an end device stays connected to a satellite as long as possible, whereas for smaller HEO and MEO constellations always connecting to the closest satellite is more beneficial.   

\section{Conclusions}
\label{sec_conclusions}

In this article we gave an overview of the design characteristics of the new private NGSO constellations proposed to operate in the Ku-, Ka-, and V-band, as specified in official applications from satellite operators to the FCC in the US. 
We subsequently discussed major challenges specific to these private constellations and also common to other (standardized) future STNs and we briefly reviewed the recent literature addressing these challenges.
Moreover, we presented the impact of two popular HO strategies on the link capacity and delay, for diverse examples of real LEO, MEO, and HEO emerging private constellations in the Ka-band. 
Our analyses showed that the optimal HO strategy depends on the constellation design, due to the very different capacity, delay, and handover rate of the different constellations. This strongly indicates that the engineering solutions for future STNs should be adjusted based on the constellation specifics, for instance by means of adaptive algorithms. Furthermore, in order to achieve global applicability and inter-working with terrestrial networks, future STN standards should be flexible and support satellite operation with the large ranges of parameter values observed in the emerging private NGSO constellations.

\section*{Acknowledgments}

Simulations were performed with computing resources granted by RWTH Aachen University under project thes0950.

\ifCLASSOPTIONcaptionsoff
  \newpage
\fi



\bibliographystyle{IEEEtran}
\bibliography{IEEEabrv, myBibliography}
\end{document}